\documentstyle[10pt]{elsart}
\begin{document}

\begin{frontmatter}
\title{Shell model studies of neutron rich nuclei\thanksref{nato}}

\thanks[nato]{Work supported in part by Nato CRG 970196,  DGES-Spain
  PB96-53 and the IN2P3.CICyT agreements.}

\author[crn]{E. Caurier}
\author[ulp]{F. Nowacki} and
\author[uam]{A. Poves}

\address[crn]{Institut de Recherches Subatomiques, IN2P3-CNRS-Universit\'e
  Louis Pasteur,
  F--67037 Strasbourg Cedex~2, France}

%\author{F. Nowacki}

\address[ulp]{Laboratoire de Physique Th\'eorique, Universit\'e
  Louis Pasteur, 3-5 rue de l'Universit\'e, F--67084 Strasbourg Cedex,
  France}

%\author{A. Poves}

\address[uam]{Departamento de F\'\i sica Te\'orica C--XI, Universidad
 Aut\'onoma de Madrid \\ 
28049 Madrid, Spain }

\maketitle
\begin{abstract}
  We discuss the present status of the description of the
  structure of the very neutron rich nuclei, in the framework of
  modern large scale shell model calculations. Particular attention is
  paid to the interaction related issues, as well as to the problems
  of the shell model approach at the neutron drip line. We present detailed
  results for nuclei around N=20  and,  more
  briefly, we discuss some salient features of the regions close to N=8,
  28  and 40. We show that most experimental features can be understood in a
  shell model context.

\end{abstract}
\end{frontmatter}

\section{Introduction}

  Our knowledge of the properties of the nuclei lying far from the
  valley of stability has increased a lot in the last decade, thanks to
  the work carried out at the isotope separators on line (ISOL) and 
  fragment separators. In some cases, mostly in light nuclei, the
  neutron and/or proton drip lines have been reached. 
  The situation will be much improved  with the
  advent of the new generation of Radiactive Ion Beam facilities that
  will be discussed at length in other papers of this volume. 
  These experimental advances have been accompanied by intense
  developments in nuclear structure theory. In the region of
  medium-light nuclei, the shell model description  in large valence
  spaces, that gives the most complete and reliable picture, has
  become available.
  A novelty in the very neutron rich side, is that the situation at
  the Fermi surface resembles that of the heavy nuclei at the valley
  of stability, with the proton and neutron Fermi levels sitting at
  different major shells. This represents a new challenge for the
  shell model approach, because two contiguous major shells have to be
  included in the valence space, in contrast with the usual
  calculations in a single major shell. This brings in new problem; on
  one side the size of the calculations can become very large,
  demanding novel shell model codes or new techniques; on the other,
  the effective in medium interaction is richer and therefore more
  difficult to keep under control. Besides, very close to the neutron
  drip line, where the physical states are only slightly bound and the
  wave functions may exhibit very long tails, the validity of an
  approach based in a Fock space representation may appear at first
  sight dubious, because of the entanglement of configuration space
  and Fock space degrees of freedom.

  The predictive power of the shell model descriptions in these new
  regions is seriously hindered by the lack of an ``universal'' shell
  model interaction. Whereas it has been demonstrated that one can
  obtain a fully reliable multipole hamiltonian from modern G-matrices,
  the monopole hamiltonian is usually incorrect \cite{duzu}. The monopole
  hamiltonian contains all the terms depending on the number of
  particles and on the isospin. The isospin dependent terms of the
  spherical mean field play a dominant role in the location of the
  different configurations far from stability and therefore determine
  which dynamical aspects will be manifest in each
  regime. Unfortunately, the experimental  knowledge of the spherical
  mean field close to the valley of stability is, in general, not
  sufficient to move safely far out. The success of the large scale
  shell model calculations depends crucially  on the correctness of
  the monopole
  hamiltonian, that's why the experimental information on some
  ``simple'' (closed shells plus or minus one nucleon) exotic nuclei
  is invaluable. 

  The first extensive survey of neutron rich nuclei in the shell model
  context is due to Wildenthal, Curtin and Brown \cite{wcb83} using
  their fitted USD interaction \cite{usd} to compute energy spectra and beta
  decay properties of all the neutron rich nuclei in the
  $sd$-shell. However, some years before, similar calculations by
  Wildenthal and Chung, when compared with the experimental data of
  $^{31}$Na, had led  these authors to entitle his paper ``The collapse of
  the shell model ordering in the very neutron rich isotopes of Na and
  Mg'' \cite{wch80}. This was a word of warning on the weird behaviour
  to be expected when exploring the far from stability land. Later, more
  experimental findings confirmed this premonitory view, and,
  nowadays, expressions like ``new phases of nuclear matter'' or 
 ``vanishing of shell closures'' are part of our current jargon.

\section{N=8: Halos}

  The neutron rich side of the $p$-shell has since long provided us
  with the most dramatic example of intruder state; the ground state
  1/2$^+$ in $^{11}$Be. The expected 0$\hbar \omega$ ``normal'' state
  lies at 300~keV. The shell model description of such inversion
  requires obviously two major shells. The mechanism of this inversion
  is common to other cases that we will study later and can be
  schematically understood as consisting of two major ingredients:

\noindent
   1) The monopole hamiltonian that  gives the
   ``unperturbed'' or spherical Hartree-Fock energy of the different
   distributions of the valence particles among the valence orbits
   (configurations). Far from stability, the energy gaps between these
   configurations  may be eroded because of the small binding of the
   orbits at the top of the well.

\noindent
   2) The multipole terms (mainly pairing and  quadrupole) that mix
   the components belonging to each configuration, 
   produce different levels of
   coherence  and different
   energy gains relative to the centro\"{\i}d
   for different configurations,
   depending on the structure of
   the spherical mean field. They can even invert the energy ordering of the
   configurations given by the monopole terms.

 There is a very nice application of this scheme to the $^{11}$Be case
 by B. A. Brown in ref.~\cite{bro94}. He examines first the $p$-$sd$
 gap evolution towards the neutron rich side, concluding that it is
 reduced, but still 4~MeV wide. Promoting a particle across the
 gap cost
 therefore  4~MeV. However, it opens the possibility for neutron
 pairing correlations (gain $\sim$~2~MeV) and also allows for the
 quadrupole coupling of the $^{10}$Be 2$^+$ with the 1d$_{5/2}$
 neutron  (gain $\sim$~2~MeV). Summing up all those contributions the
 intruder wins. Notice the subtle balance between spherical mean field
 properties and correlations, the latter depending very much on the
 detailed location of the orbits around the Fermi level. When many
 particles many holes excitations are at play, the monopole effects
 are more involved, as discussed by A. Zuker \cite{zuk98} in the case
 of the 4p-4h 0$^+$ excited state of  $^{16}$O at $\sim$~6~MeV and
 clustering (mainly $\alpha$ correlations) effects are surely present in
 the physical solution.
 
 These calculations and a few similar ones \cite{n8,bor98}, have provided a
 solid shell model interpretation to the behaviour of the nuclei in
 the region, in agreement with or confirmed by subsequent experiments.
 A most prominent member of this region is  $^{11}$Li, which sits at
 the drip line and has a very small ($\sim$~200~keV) two-neutron
 separation energy. The last two neutrons have a very large spatial
 extension, forming what is called a neutron halo \cite{halo}.
 In spite of the
 exotic matter distribution and of the closeness of the continuum, the
 shell model picture can  still cope with many of the structural
 properties of $^{11}$Li. The calculations predicted a ground state of
 $^{11}$Li dominated by a configuration with two neutrons in the 
 2s$_{1/2}$ orbit. Indeed, the most recent experimental information
 confirms this extreme \cite{bor99}. Another fingerprint of the
 dominance of intruder configurations in $^{11}$Li was provided by a
 classical beta decay experiment, measuring its  lifetime and the
 branching ratio of its decay to the first excited 1/2$^{-}$ state
 in $^{11}$Be. The results were only compatible, by large, with the
 assumption of $s$-wave dominance \cite{bor98}.
 Very recent experimental work at MSU, has shown that a similar
 situation happens in $^{12}$Be. The expectedly semi-magic isotope of
 Berilium (N=8) turns out to be dominated by the intruder
 configuration with two neutrons in the $sd$-shell \cite{navin}. With
 this, it joins its forerunner cousin $^{32}$Mg in the realm of the
 intruders.

\section{N=20:   Intruders}

It was in this region were the massive breaking of a semi-magic closure
far from stability was first detected, in what is known as ``the
island of inversion'' around $^{31}$Na. The anomalous experimental
data on the mass and the spin of $^{31}$Na
\cite{na31} were attributed to a transition from
spherical to prolate shape at N=20 \cite{campi}.  Later, the measures were
extended to other Neon, Sodium and Magnesium isotopes 
\cite{n20exp} and were interpreted in a shell model context as due to the
inversion of the neutron closed shell configuration and a 2p-2h
intruder configuration, intrinsically deformed \cite{pr87}. The
intrusion mechanism, that works as we have explained in section 2, was
already sketched in ref.~\cite{pr87}. The N=20 quasiparticle gap
diminishes when the neutron rich area is approached. For instance, its
experimental value for $^{40}$Ca is $g$=7~MeV, for $^{36}$S is
$g$=~5.6~MeV and for $^{34}$Si is $g$=~5.1~MeV. $^{34}$Si is a good
reference because it is clearly semi-magic, as we shall discuss later,
and very neutron rich. Extrapolating smoothly these numbers one should
expect $g\sim$~4.6~MeV for $^{32}$Mg, whereas experimentally
$g$=~3.6~MeV. This difference may be actually due to the  onset of
deformation.  A two particle-two hole neutron excitation across the
N=20 closure, would accordingly cost $\sim$~7~MeV, much less than what
it takes in $^{40}$Ca. When the calculations are performed, it turns
out that the gain in correlation energy of the intruder state
overshoots the monopole gap by 1.5~MeV, producing the famous
inversion. Most of the intruder's gain in correlation energy is
quadrupole, because of the presence of open shell $sd$ protons and
open shell $pf$ neutrons. When we move to $^{34}$Si, the gap becomes
larger and the quadrupole correlation is hindered due to the closure of
the 1d$_{5/2}$ proton subshell. As a result, the neutron closed shell
is now the ground state and the intruder becomes an excited state.

 Different groups have  made calculations in this region. 
 In ref.~\cite{bro.msu} the diagonalizations were supplemented with a
 weak coupling approximation to delineate the contour of the ``island
 of inversion''. More recently \cite{fuku,siis}
 similar calculations have been undertaken using
 different truncations of the $sd$-$pf$ space. In ref.~\cite{us98} the
 full $sd$-shell for neutrons and the full $pf$-shell for neutrons is
 considered. The  calculations allow up to two particle jumps from
 the $sd$-shell to the $pf$-shell. Besides, it turns out that the
 cross shell proton
 excitations are irrelevant in this zone. There is
 a good level of agreement between the results of  these groups, in
 particular they share the same strong and weak points. One of the
 strong points is the description of the structure of the isolated
 intruders as well as its location relative to the normal states for
 nuclei N=20. Another common strong point concerns the predicted
 limits in Z of the island of inversion; in all the calculations 
 only Ne, Na and
 Mg belong to it, while F and Al sit at its very edge.
 In the weak side are the limits in N of the
 region. Different choices of the monopole hamiltonian produce small
 shifts in the borders. As a consequence, N=19 and N=22 are inside or
 outside depending on the calculation. Also in the weak side is the
 amount of mixing between normal and intruder states. 
  Very recently, the Quantum
 Monte Carlo diagonalization method has been also implemented in this
 region, using as valence space the $sd$-shell plus the two lower
 $pf$-shell orbits, 1f$_{7/2}$ and 2p$_{3/2}$ \cite{otsu99}. The
 effective interaction employed in this reference has been adjusted to
 produce a  1d$_{3/2}$--$fp$ gap that decreases rapidly between
 $^{34}$Si ($g$=4.4~MeV) and $^{28}$O ($g$=1.2~MeV), whereas in our
 case these figures are 4.7~MeV and 3.4~MeV respectively. It is
 evident that this choice results in an enhancement of the intruder
 mixing and in an enlarged ``island of inversion''.
 
  We shall now present some of our latest results. We use the same 
  valence space we had in ref~\cite{us98} and essentially the same
  effective interaction. A modification has been forced by the recent
  experimental measure at Isolde \cite{guy} of the excitation energy
  of the 3/2$^-$ state in  $^{35}$Si (1~MeV). In order to fix the monopole
  terms of the cross shell interaction in the $sd$-$pf$ valence space
  this information is vital. In our old interaction we had taken the
  conservative view and had put the 3/2$^-$ state at 2~MeV (as in
  $^{41}$Ca). The effect of this change on  the results of
  ref.~\cite{us98} is not dramatic  and amounts  to
  enhance  moderately the quadrupole correlations,  increasing the
  binding energy and the deformation of the intruders. Besides, this
  modification binds $^{31}$F, in agreement with the most recent
  experimental result \cite{saku}. In what follows we 
  compare  the structure of the normal and
  intruder states in the different nuclei of interest and give  our
  predictions for their relative position.

\begin{table}
\caption{Properties of the even magnesium isotopes. N is for normal
  and I for intruder. Energies in MeV, BE2's in
  e$^2$fm$^4$ and Q's in efm$^2$ in all the tables}
\vspace{0.1cm}
\begin{tabular}{cccccccccc}
\hline 

   &  & $^{30}$Mg  &   &   & $^{32}$Mg   &  &     & $^{34}$Mg   &  \\
\hline
    & N  & I  & EXP  &  N  & I  & EXP   & N  & I  & EXP \\
\hline

$\Delta$E(0$^+_{\textrm{I}}$) &      & +3.1  &     &    & -1.4  &   
&    & +1.1   & \\

0$^+$ &    0.0 & 0.0  &      & 0.0 & 0.0  &   
& 0.0  & 0.0   & \\

2$^+$     &1.69   & 0.88  &1.48      &1.69 & 0.93  & 0.89  
  &1.09  & 0.66   &(0.67) \\
4$^+$     &4.01   & 2.27  &      &2.93 & 2.33  & (2.29)  
  &2.41  & 1.86   &(2.13) \\
6$^+$     &6.82   &3.75   &      &9.98   &3.81   &    
  &3.52  & 3.50   & \\
BE2     &   &   &   &   &   &   &   &      & \\
2$^+$ $\rightarrow$ 0$^+$  &   53   &112   & 59(5)     &36 & 98
  & 90(16)    &75  & 131   & \\
4$^+$ $\rightarrow$ 2$^+$  &   35   &144   &      &17   &123   
  &  &88   & 175   & \\
6$^+$ $\rightarrow$ 4$^+$  &   23   &140   &      &2   &115   &   
  &76  & 176    & \\
Q$_{spec}$(2$^+$)  &    -12.4  &-19.9   &      & -11.4  &-18.1   &
  &-15.4  & -22.7   & \\
\hline
\end{tabular}
\end{table}

  We start the tour with the even Mg isotopes with N=18, N=20 and
  N=22. Our results are gathered in table 1. In $^{30}$Mg the
  configuration with normal filling gives clearly the best
  reproduction of the (scarce) existing experimental data \cite{pry}.
  Notice 
  however, that the 2p-4h intruder is very collective, actually
  as much  as the intruder in $^{32}$Mg. We can
  therefore conclude that $^{30}$Mg is outside the inversion
  zone. In $^{32}$Mg the
  situation is the opposite. Our calculation places the 2p-2h intruder
  well below the closed shell configuration. The differences between
  both are manifest, and the excellent agreement between the
  properties of the
  calculated intruder and the experimental data (the 2$^+$ excitation
  energy \cite{guille}, the  0$^+$~$\rightarrow$~2$^+$ BE2
  \cite{moto} and the 4$^{+}$ excitation energy \cite{aza}) make it posible to
  assign this configuration unambiguously. Data and calculations
  suggest, at low spin, a prolate deformed structure, certainly perturbed, with
  $\beta \sim$~0.5. But, what about the mixing between different np-nh
  configurations? Indeed, the true physical state must be 
  mixed  to a larger or smaller extent. However, in view of the
  excellent agreement of our fixed 2p-2h solution with the experiment,
  it is clear that any mixing would deteriorate it. The way out
  of this dilemma --to mix or not to mix-- refers to the
  effective  interaction. The $sd$-shell and $pf$-shell parts of our
  effective interaction are well suited for 0$\hbar \omega$
  calculations, and contain implicitly part of the effects of the cross shell
  mixing. Thus, in a mixed calculation one has to take care
  of properly unrenormalising the interaction. With this caveat, the mixed
  results may come back to agree with the experiment.  
  In $^{34}$Mg the normal configuration that contains
  two $pf$ neutrons, is already quite collective. It can be seen in
  the table that it resembles very much the intruder configuration in
  $^{32}$Mg. Its own 4p-2h intruder is even more deformed ($\beta
  \sim$~0.6) and a better rotor. Therefore, it is more difficult to
  make a sharp distinction between them both and their different
  mixed combinations. 
  In our calculation the normal state is
  1~MeV below the intruder. However, it is by no means  excluded that
  they could be much closer or even than the intruder would come below. We
  have put in parenthesis the very recent and preliminary results
  from Riken \cite{mg34} that seem to favour the intruder option. Let's
  mention however, that results equivalent to those given by the
  intruder configuration alone, can be also obtained with a 50\% mixed
  solution, provided the $pf$-shell pairing is reduced. In the QMCD
  calculations of ref.~\cite{otsu99}, the ground state band is
  dominantly 4p-2h; the 2$^+$ comes at the right place but the 4$^{+}$
  is too high, making the solution to over-rotate. Clearly, $^{34}$Mg
  is at the edge of the ``island of inversion'', whether it is more on
  the inside or the
  outside is (theoretically) a matter of subtle arrangements that can
  only be decided  by better experimental data.

\begin{table}
\caption{Properties of the even neon isotopes. N is for normal
  and I for intruder.}
\vspace{0.1cm}
\begin{tabular}{cccccccccc}
\hline

  & &   $^{28}$Ne  &    &  & $^{30}$Ne   &  &     & $^{32}$Ne   &  \\
\hline
  &   N  & I  & EXP  & N  & I  & EXP  &  N  & I  & EXP \\
\hline

$\Delta$E(0$^+_{\textrm{I}}$) &      & +2.6  &   &       & -1.4  &   &
    & +1.5   & \\

0$^+$ &    0.0 & 0.0     &   & 0.0 & 0.0     &
& 0.0  & 0.0   & \\

2$^+$  &   1.81   & 0.87  &(1.32)    &1.90 & 0.85     &
  &1.01  & 0.68   &  \\
4$^+$  &   3.34   & 2.21  &     &2.87 & 2.08  &   
  &2.09  & 1.82   &  \\
6$^+$     &6.35   &3.90      &   &    &3.61   &    
  &3.29  & 3.42   & \\
BE2     &   &   &   &   &   &  &  &    & \\
2$^+$ $\rightarrow$ 0$^+$     &36   &78   & 54(27)     &29 & 72
  &     &72  & 100   & \\
4$^+$ $\rightarrow$ 2$^+$     &31   &105   &     &22   &97   &
    &74   & 137   & \\
6$^+$ $\rightarrow$ 4$^+$     &15   &104   &      &   &89   &   
  &60  & 133    & \\
Q$_{spec}$(2$^+$)  &    -1.2  &-17.8   &   &    -1.1  &-16.4   &
   &-13.7  & -20.0   & \\
\hline
\end{tabular}
\end{table}

In table 2 we have collected the results for the even Neon isotopes to
which, {\it mutatis mutandis}, most of the arguments advanced in the
discussion of the magnesiums apply. Now, the experimental information
is even meagrer than before. Let's just comment on the $^{28}$Ne case
because it has been argued in ref.~\cite{otsu99} that it could be
substantially more mixed than its neighbour $^{30}$Mg. This claim
originates in the comparison between the $sd$ prediction for the 2$^+$
excitation energy (1.81~MeV using the USD interaction) and the
experimental result (1.32~MeV \cite{aza}). However, there could be
another explanation; that the discrepancy were due, instead, to a
defect of the USD interaction, for the experimental 2$^+$ excitation
energy of $^{28}$Ne (1.32~MeV) is only slightly lower than that of
$^{30}$Mg, (1.48~MeV). The recent measure of the BE2 \cite{pry} does
not  settle the case yet, because its large error bar do not discards
a low mixing scenario.
  In the N=19 isotones $^{29}$Ne and  $^{31}$Mg the situation is even
  more complex, because the normal configurations are almost
  degenerate with the opposite parity 1p-2h intruders, while the 2p-3h
  intruders appear a bit above. The competition for the ground state
  is between 3/2$^+$  and 3/2$^-$ in both cases. The beta decay
  data from Isolde \cite{klotz} favour the positive parity for the
  ground state of  $^{31}$Mg.  On its side, the calculation
  explains the occurrence of  such a high level density at low
  excitation energy in the experiment. 
  In the N=21 isotones $^{31}$Ne and  $^{33}$Mg the lowest
  configurations are the positive parity 2p-1h intruders and the
  negative parity 3p-2h intruders. Like in the N=19 case they are
  nearly  degenerate and the ground state candidates are  3/2$^+$ and
  3/2$^-$. In a recent experimental study of the decay of $^{33}$Na
  \cite{walter} it is found that the ground state is 3/2$^+$ . This
  agrees with our calculation that, in addition,  predicts  a 3/2$^-$
  at 300~keV.
  In table 3. we show the results for $^{31}$Na. In this nucleus the
  intruder configuration is clearly dominant and very distinct from
  the normal one. The calculation reproduces the occurrence of 3/2$^+$
  as spin of the ground state as well as the excitation energy of the
  5/2$^+$ state, recently measured at MSU \cite{pryt}. In this
  reference, shell model calculations  along the same lines than ours,
  although in a somewhat smaller valence space, are reported. They 
  agree reasonably well with the present results.

\begin{table}
\caption{$^{31}$Na level scheme. N is for normal
  and I for intruder.}
\vspace{0.1cm}
\begin{tabular}{ccccccccc}
\hline

 \multicolumn{9}{c}{$\Delta$E(intruder$-$normal)=$-$1.6} \\[0.2cm]
\hline
& 2J  & N &  & 2J &  I(2p-2h) & & 2J& exp   \\
\hline
 &5$^+$ & 0.0  &  & 3$^+$   & 0.0 &   & 3$^+$  & 0.0  \\ 
 &3$^+$ & 0.45  &  & 5$^+$   & 0.28 &   & (5$^+$)  & 0.35(2)  \\
 &1$^+$ & 3.18  &  & 7$^+$   & 1.06 &   &   &   \\
 &7$^+$ & 4.42  &  & 1$^+$   & 2.28 &   &   &   \\
\hline

BE2(5$^+ \rightarrow$ &3$^+$) & 62  &BE2(3$^+ \rightarrow$  & 5$^+$)
    & 216 &   &   & \\
    &1$^+$) & 12  &  & 7$^+$)   & 110 &   &  &   \\
    &7$^+$) & 40  &  & 1$^+$)   &   6 &   &  &   \\
\hline 
\end{tabular}
\end{table}

\begin{table}
\caption{$^{34}$Si level scheme. N is for normal
  and I for intruder.}
\vspace{0.1cm}
\begin{tabular}{ccccccccccc}
\hline 

  & N &exp & &  & I(2p-2h) &exp &  &  &I(1p-1h) & exp   \\
\hline

0$^+$ & 0.0  & 0.0 &    & 0$^+$ & 1.7 & (2.1) &   & 4$^-$  
& 4.19  & 4.38  \\

2$^+$ & 4.86  & 5.3 &   & 2$^+$ & 3.0 & 3.3 &   & 3$^-$  
& 4.40  & 4.26  \\

4$^+$ & 7.92  &   &    & 4$^+$ & 4.7 &   &   & 5$^-$  
& 4.53  & 4.97  \\
\hline 
\end{tabular}
\end{table}

Finally, in table 4. we move outside of the ``island of inversion''. In
its ``normal'' ground state $^{34}$Si is closer to doubly-magic than to
semi-magic. The different intruders fit
very nicely with the experimental data. Notice the not very frequent
level scheme, with a 0$^+$ as first excited state. It stems from the
calculation that this state is oblate with $\beta \sim$~0.4. The band
built upon it is not very regular, but the first couple of transitions
are in accord with the rotor picture and are consistent with the value
of the 2$^+$ spectroscopic moment.

\section{N=28: Coexistence}

 If we move towards the neutron drip line, the Mg isotopes continue being 
 deformed; even the N=28, $^{40}$Mg, which sits at the drip line.
 Another candidate to
 shell closure disappearance! Experimentally N=28 has been reached for
 Z as low as 14 and there is a lot of debate on the physical
 interpretation of the data. In ref.~\cite{reta} we studied the
 region. When the reduction of the $1f_{7/2}-2p_{3/2}$ splitting in
 $^{35}$Si is plugged in the calculations, its  main effect is to erode
 the N=28 shell closure, bringing the 2$^+$ excitation energy and the
 BE2(0$^+$ $\rightarrow$  2$^+$) in  $^{44}$S and  $^{46}$Ar into full
 agreement with the data \cite{glas}. In  $^{42}$Si the
 excitation energy of the first 2$^+$ drops by $\sim$~1~MeV. We can
 follow the behaviour of the sulphur isotopes in the same valence
 space crossing  two magic numbers and fifteen units of mass.
 $^{36}$S is spherical, its N=20 neutron closure is reinforced by the
 proton  2s$_{1/2}$ closure, resulting in a 2$^+$ at 3.5~MeV. Adding
 neutrons, the 2s$_{1/2}$ and 1d$_{3/2}$ become degenerate and, at
 N=26,  $^{42}$S is a nearly perfect prolate rotor. At N=28 the
 spherical and the deformed solutions appear at the same energy
 and mix at 50\%.
  $^{43}$S could provide a nice example of shape coexistence and
 isomerism. Our calculation produces a 3/2$^-$ deformed ground state, a
 low lying 7/2$^-$ spherical isomer and another 7/2$^-$ at higher
 energy, belonging to the ground state band. All these in full
 correspondence with the experimental results of ref.~\cite{sara}.
  One could even attempt N=40, but  realism advises to increase Z a few
 units. We suggested  a few years ago that $^{64}$Cr could be another
 ``semi-magic'' prolate rotor, because of the strong quadrupole
 coherence of the intruder configurations with four $pf$ protons and four
 $gds$ neutrons. Preliminary evidence of such  behaviour has been
 reported in its neighbour  $^{66}$Fe \cite{hanna}. For the moment no
 signs of weakening of N=50 in $^{78}$Ni have been found \cite{daug}.

%\section{Conclusion}

\end{document}